\documentclass[aps,floatfix,twocolumn,showpacs,prl]{revtex4}%
\usepackage{amssymb}
\usepackage{amsmath}
\usepackage{bm}
\usepackage{epsfig}
\usepackage{comment}
\usepackage{graphicx}
\newcommand{\be}{\begin{eqnarray}} 
\newcommand{\ee}{\end{eqnarray}} 
\newcommand{\bml}{\begin{multline}}
\newcommand{\eml}{\end{multline}}

\begin{document}
\title[]{Symmetry relations for spin-resolved exchange correlation kernels
 \\in soft magnetic layered systems}
\author{Ina Yeo}
%\email{}
%\thanks{Fax: +82-2-554-1643}
\author{Kyung-Soo Yi}
\affiliation{Department of Physics
and
Research Center for Dielectric and Advanced Matter Physics\\
Pusan National University, Busan 609-735, Korea}
\date{\today}

\begin{abstract}
We first exploit the physical condition satisfying the symmetry relation
of the ``exact'' spin-resolved exchange correlation kernel
based on the ``mixed scheme'' in soft magnetic layered systems.
The conditions are derived and examined by
taking into account the field gradients of the magnetic moment as well as that of the electric moment.
We also exploit the physical condition by means of deviation distribution function
suitable for complex electronic structures with arbitrary $\zeta$.
\end{abstract}


%\keywords{Exchange correlation kernel, 
%Spin symmetry relation, Electrochemical potential, Spin injection}

\maketitle
Density functional theory (DFT) has been a chief support of the condensed matter theory for its 
practical prediction of various properties of many-electron system.
It has resolved many properties such as 
the ground state energies, forces on the atoms or molecules, and their equilibrium 
positions \cite{Giuliani}. Its local density approximations including density gradient corrections
have generally been used for their simplicity
and accuracy \cite{Giuliani,phil}.
Unfortunately, there have been some exceptions for their accuracy.
On top of that, semilocal approximations fail to describe the long range interaction between electrons
whose densities do not or weakly overlap.
As an alternative,
the ``mixed scheme'' has been proposed \cite{julien,pazini,fran,Gori,Gonis}.
In this scheme, the Coulomb electron-electron interaction is decomposed into the short range (SR) part and
the complementary long range (LR) part.
The key concept is then to
 use a semilocal approximation for SR part and a more appropriate many-body methods
for LR part. %based on the idea to improve the description of the dispersion 
%interaction between distant electrons. 

In view of practical interests for spin transport properties of soft magnetic layered systems,
the resolution of topics of the weak dispersion or van der Waals forces 
is clearly important. 
Additionally, there has been growing great deal of interests
for the magnetic information storage system basically containing soft ferromagnets \cite{apal,hei,Rebei,tse}.
These storage applications are
based on the idea of the ``spin-torque switching'' in ferromagnetic-normal-ferromagnetic metal hybrid structures. 
The idea has been proposed by
Slonczewski \cite{Slonc1,Slonc2} and Berger \cite{Ber} as an attractive alternative of controlling the magnetization
in place of the traditional way of modulating magnetic fields.
Considerable evidence for the concept has been accumulated by demonstrating
that the magnetization can be switched or precessed in a soft ferromagnet, because a current 
exerts ``spin-transfer torque'' on a thin free magnetic layer, which is polarized in a thick ``fixed'' magnetic layer
\cite{apal,hei,Rebei,tse,Slonc1,Slonc2,Ber}.
Such interests inspired us to propose symmetry relations for the exchange correlation kernel (XCK)
in soft magnetic systems involving intricate boundary conditions 
 and corresponding spin scatterings. 

As one of the central signatures in DFT,
XCK can be useful for giving direct insight 
into the dynamics of spin interactions in multilayer system.  
XCK is a basic
concept in describing many body correlation effects in an inhomogeneous electron liquid
and is a vital element of the spin susceptibility.
Spin-resolved XCK $f^{\rm xc}_{ss'}(\zeta)$ is defined by 
\begin{eqnarray}
f^{\rm xc}_{ss'}(\zeta)\equiv
\frac{\partial^2 E^{\rm xc}(\boldsymbol r,\boldsymbol r';\zeta)}
{\partial n_{s'}(\boldsymbol r)\partial n_{s}(\boldsymbol r')}\label{first}
\end{eqnarray}
where $E^{\rm xc}$ is the total XC energy functional in many electron system with spin polarization $\zeta$.
Here, $n_{s}(n_{s'})$ denotes the electron density in spin state $s(s')$.
$f^{\rm xc}_{ss'}(\zeta)$ satisfies the symmetry relation 
\begin{eqnarray}f^{\rm xc}_{s\bar s}(\zeta)=f^{\rm xc}_{\bar s s}(-\zeta)\end{eqnarray}
which is a key property of XCK and equally a fundamental idea of importance in condensed matter. 
The specific symmetry relation \begin{eqnarray}f^{\rm xc}_{s\bar s}(\zeta)=f^{\rm xc}_{\bar s s}(\zeta)\end{eqnarray}
also plays a significant role to investigate the situations in artificial composite structures \cite{ina}. 
 However, prior to our previous work, there was no attempt to interpret XCK in terms of directly measurable quantities
such as spin current $I_{s(\bar s)}$.
 Although considerable studies have been devoted to investigate $I_{s(\bar s)}$,
 theoretical approaches have substantially been limited to oversimplified multilayer systems
 unsuitable for more realistic applications.
  
  %\propto-\frac{\partial\mu_{s(\bar s)}}{\partial
%x}$ where $\mu_{s(\bar s)}$ is the electrochemical potential in spintronics

 The introduced key idea in this letter is the accurate relations guaranteeing the specific symmetry relation
in soft magnetic layered systems. The accurate relations are derived explicitly by 
including not only the field gradient of the electric moment but also that of the magnetic moment.
We also explicitly show the relation which guarantees the specific symmetry relation 
  by using the solution of Boltzmann equation without neglecting anisotropic terms in systems of various dimensions.
The condition satisfying the specific symmetry relation could be 
related to the spin density variation $\nabla n_{s(\bar s)}$ and then 
$\nabla n_{s(\bar s)}$ to the variation of electrochemical potential $\nabla \mu_{s(\bar s)}$.  
Hence a general inhomogeneous multilayer system
with arbitrary magnetization alignments can be explored easily 
by examining properties of XCK on the basis of $I_{s(\bar s)}\propto-\frac{\partial\mu_{s(\bar s)}}{\partial
x}$.
 
  Density functional based on the ``mixed scheme'' is given by the sum
of the kinetic energy $T$ of a noninteracting particle system, SR and LR potential energies
$U(\equiv U_{\rm S}+U_{\rm L})$, 
and unknown SR and LR functionals $E^{\rm xc}(\equiv E^{\rm xc}_{\rm S}+E^{\rm xc}_{\rm L})$;
$
E[n(\boldsymbol r)]=T[n(\boldsymbol r)]+U[n(\boldsymbol r)]+E^{\rm xc}\nonumber
$
in terms of one particle density in N-electron system
\[n(\boldsymbol r)=N\sum_{s_2,\cdots,s_N}\int\lvert\Psi(\boldsymbol r s,\boldsymbol r_2 s_2\cdots,\boldsymbol r_N s_N)\lvert^2
d\boldsymbol r_2,\cdots,d\boldsymbol r_N.\] 
Once SR (LR) part is given by $T[n(\boldsymbol r)]+U_{\rm S(L)}[n(\boldsymbol r)]+E^{\rm xc}_{\rm S}$,
the complementary LR (SR) part is the difference between the universal functional $E[n(\boldsymbol r)]$ 
and the SR (LR) part.
In pair density theory giving a more improved ground state energy 
than in one particle density \cite{Gonis}, 
the spin-summed pair density is given by
$n(\boldsymbol r_1,\boldsymbol r_2)=\frac{N(N-1)}{2}\sum_{s_1 s_2}\gamma_{s_1 s_2}(\boldsymbol r_1,\boldsymbol r_2)$
where $\gamma_{s_1 s_2}(\boldsymbol r_1,\boldsymbol r_2)$ is the spin-resolved diagonal 
of the two-body reduced density matrix \cite{Gori}.
%\begin{eqnarray}
%\gamma_{s_1 s_2}(\boldsymbol r_1,\boldsymbol r_2)=\sum_{s_3,\cdots,s_N}
%\int\lvert\Psi(\boldsymbol r_1 s_1,\cdots,\boldsymbol r_N s_N)\lvert^2d\boldsymbol r_3,\cdots,d\boldsymbol r_N.\nonumber
%\end{eqnarray}
%Any extended accurate approximation to spin density $\rho(r)$ is available to get exact density functional. 
The exact energy density functional is given by $E[n(\vec x)]=T[n(\vec x)]+U$ with
$\vec x=(\boldsymbol r_{1},\boldsymbol r_{2})$.
Here, $U(\equiv U_{\rm S}+U_{\rm L})$ consists of the external potential of a given pair and 
the interaction potential between particles forming two pairs at 
$\vec x_i=(\boldsymbol r_{i1}, \boldsymbol r_{i2})$ 
and $\vec x_j=(\boldsymbol r_{j1},\boldsymbol r_{j2})$ \cite{Gonis}.
Hence, the total XC energy functional $E^{\rm xc}$ in an interacting system
can be described in terms of pair density $n(\vec x)$
\begin{eqnarray}
E^{\rm xc}[n(\boldsymbol r)]=T[n(\vec x)]-T[n(\boldsymbol r)]+U[n(\vec x)]-U[n(\boldsymbol r)].\nonumber
\end{eqnarray} 
 $E^{\rm xc}$ contains the unknown remaining kinetic term 
$\Delta T\equiv T[n(\vec x)]-T[n(\boldsymbol r)]$ as well as the interparticle interaction potential.
Here, $\Delta T$ denotes the difference between the exact kinetic term $T[n(\vec x)]$ in an interacting scheme
and noninteracting counter part $T[n(\boldsymbol r)]$.
$f^{\rm xc}_{s\bar s}$,  
the second derivatives of $E^{\rm xc}$ with respect to spin densities $n_{s}(\boldsymbol r)$ 
and $n_{\bar s}(\boldsymbol r')$
at a given pair position $\vec x_i=(\boldsymbol r,\boldsymbol r')$,
is now written in the form
\begin{eqnarray}
f^{\rm xc}_{s\bar s}(\zeta)
=\frac{\nabla_{\boldsymbol r}\nabla_{\boldsymbol r'}[\Delta T[n(\vec x)]+U[n(\vec x)]-U[n(\boldsymbol r)]]}
{\nabla n_{\bar s}(\boldsymbol r) \nabla n_{s}(\boldsymbol r')}.
\end{eqnarray}
For the defined XCK by Eq.(\ref{first}), the symmetry relation  
$f^{\rm xc}_{s\bar s}(\zeta)=f^{\rm xc}_{\bar s s}(-\zeta)$ 
is trivial in an inhomogeneous system of broken spin symmetry
since
majority and minority spins interchange their orientations and positions 
with the reversed polarization $-\zeta$.
That is, based on the assumption that the physical condition
\begin{eqnarray}
\frac{\nabla n_{\bar s}(\boldsymbol r')}{\nabla n_{s}(\boldsymbol r')}=
\frac{\nabla n_{\bar s}(\boldsymbol r)}{\nabla n_{s}(\boldsymbol r)}\label{key}
\end{eqnarray}
is fulfilled (i.e., the ratios of spin density gradients
are the same at two different positions $\boldsymbol r$ and $\boldsymbol r'$),
the specific symmetry relation 
\begin{eqnarray}
f^{\rm xc}_{s\bar s}(\zeta)=\frac{\partial^2 E^{\rm xc}(\boldsymbol r,\boldsymbol r';\zeta)}
{\partial n_{\bar s}(\boldsymbol r)\partial n_{s}(\boldsymbol r')}
=\frac{\partial^2 E^{\rm xc}(\boldsymbol r,\boldsymbol r';\zeta)}
{\partial n_{s}(\boldsymbol r)\partial n_{\bar s}(\boldsymbol r')}=f^{\rm xc}_{\bar s s}(\zeta)\label{syr}
\end{eqnarray}
is obtained trivially.
So to speak, we explicitly show the specific symmetry relation by using XCK defined on the ``mixed scheme''.
Hence it is elucidated that Eqs.(\ref{key}) and (\ref{syr}) can be used to investigate soft magnetic systems 
without difficulty. 
Equation(\ref{key}) thus can be useful to examine the adequacy of the specific symmetry relation
in particular for the present weakly interacting systems.

The continuity equation for the probability density
$\rho_{\boldsymbol M}(t)[\equiv\mu_{\rm B}(n_{s}-n_{\bar s})]$
is written in form \cite{brown}
\begin{equation}
\frac{\partial\rho_{\boldsymbol M}(t)}{\partial t}=-\nabla\cdot\boldsymbol j_{\boldsymbol M}(t)
\end{equation}
where the probability current density $\boldsymbol j_{\boldsymbol M}(t)[\equiv\mu_{\rm B}
(\boldsymbol j_{s}-\boldsymbol j_{\bar s})]$
on $\boldsymbol M$-sphere is given by \cite{apal}
\begin{equation}
 \boldsymbol j_{\boldsymbol M}(t)=\rho_{\boldsymbol M}(t)\dot{\boldsymbol M}_{\rm det}
 -D\nabla\rho_{\boldsymbol M}(t)\label{bro}
\end{equation}
where $D$ is the diffusion constant in the case that diffusion constants of up-spin and down-spin are the same. 
Here, $\dot{\boldsymbol M}_{\rm det}$ is the deterministic part of the Landau-Lifshitz (LL)
equation for the evolution of a uniform magnetization $\boldsymbol M(t)$.
The deterministic part is the sum of the conservative precession, dissipative LL damping, 
Slonczewski current-induced term \cite{apal}, and a neglected term in the current-induced part \cite{tse};
\begin{eqnarray}
\dot{\boldsymbol M}_{\rm det}=&&-\gamma\boldsymbol M\times\boldsymbol H_{\rm cons}
-\gamma\alpha M\hat{m}\times(\hat{m}\times
\boldsymbol H_{\rm cons})\nonumber\\
&&-\gamma JM\hat{m}\times(\hat{m}\times\hat{m}_{\rm p})-\frac{\gamma\hbar\tilde{g}^{s\bar s}}
{8\pi M} \hat{m}\times\frac{d\hat{m}}{dt}\label{total}
\end{eqnarray}
where $\gamma$, $M$, $\alpha$, and $J$  are, respectively, 
the gyromagnetic ratio, saturation magnetization, LL damping constant, and 
an empirical constant with units of magnetic field which is proportional to the current.
Here, $\hat{m}\equiv{\boldsymbol M}/M$ and $\hat{m}_{\rm p}$ is the unit vector along the magnetization and the easy axis.
In the first and second terms, $\boldsymbol H_{\rm cons}$ is the field around which 
$\boldsymbol M$ precesses as conservative and
is related with the magnetic energy $E_{\boldsymbol M}$
via $\mu_{0}\boldsymbol H_{\rm cons}=-\frac{\partial E_{\boldsymbol M}}{\partial\boldsymbol m}$. 
The magnetic energy $E_{\boldsymbol M}$ is the 
sum of the anisotropy energy and the coupling term with an external field \cite{apal}.
When the last term (containing the mixing conductance $\tilde{g}^{s\bar s}$) is neglected,
 the current-induced part in Eq.(\ref{total}) is consistent with the result of Slonczewski \cite{tse}. 

 The current for up (down)-spin carriers can be written by
\begin{eqnarray}
\boldsymbol j_{s(\bar s)}&&= -n_{s(\bar s)}\nu\boldsymbol E\pm n_{s(\bar s)}
\frac{\dot{\boldsymbol M}_{\rm det}}{2}-D_{s(\bar s)}\nabla n_{s(\bar s)}\label{gecu}
\end{eqnarray}
where $\nu$ is the mobility and $D_{s(\bar s)}$ is the spin-up (down) diffusion constant.
Equation(\ref{gecu}) is
based on the fact that the drift velocity is proportional to the force on the conduction electron.
 The force originates from the field gradients of its electric and magnetic moment 
(i.e., $\kappa\boldsymbol v_{s}=\boldsymbol F_{s}=-e\boldsymbol E\pm\mu_{\rm B}\nabla H_{\rm eff}$) \cite{hei}.
Here, the effective field $H_{\rm eff}$ includes the external field and the molecular field coupled to the 
conduction electron via an exchange constant.
Hence, the spin-up (down) velocity $v_{s(\bar s)}=-\nu\boldsymbol E
\pm\dot{\boldsymbol M}_{\rm det}/2$ is applied in Eq.(\ref{gecu}),
because
the difference between the velocities of up-spin and down-spin 
is given by $v_s-v_{\bar s}=2\mu_{\rm B}\nabla H_{\rm eff}/\kappa=\dot{\boldsymbol M}_{\rm det}$
from Eq.(\ref{bro})

The probability density $\rho_{\boldsymbol M}$
often depends only on energy, for instance, in a thermal equilibrium or
a nonequilibrium steady state without LL damping $\alpha$ and current $J$.
%As long as $\alpha$ and $J$ are small, the variation of $\rho$ along a constant energy orbit%
%is the first order of $\alpha$ and $J$ \cite{apal}.
For this reason, the density $\rho_{\boldsymbol M}(t)$ 
can be replaced with $\rho_{\boldsymbol M}(\varepsilon,t)$ from the assumption that $\rho_{\boldsymbol M}(\varepsilon,t)$
depends only on energy. 
Still energy dependence can be different in distinct positions of sphere. 
 Together with  $\rho_{\boldsymbol M}(\varepsilon,t)$, the spin-up (down) probability current density can be obtained by
\begin{eqnarray}
&&j_{s(\bar s)}(\varepsilon,t)=\oint[\boldsymbol j_{s(\bar s)}\times d\boldsymbol M]\cdot\hat{m}\nonumber\\
&&=n_{s(\bar s)}(\varepsilon,t)\Big\{\nu I^{\rm EE}\mp\frac{\gamma}{2}[
(\alpha M-\frac{\gamma\hbar\tilde{g}^{s\bar s}}
{8\pi})I^{\rm HE}-J\hat{m}_{p}\cdot I^{\rm M}]\Big\}\nonumber\\
&&-D_{s(\bar s)}\frac{\partial n_{s(\bar s)}(\varepsilon,t)}
{\partial \varepsilon}\mu_{0}I^{\rm HE}\label{curr}.
\end{eqnarray}
%$j_{s(\bar s)}(E,t)$ contains a damping term with $I^{\rm HE}$ and
%a Slonczewski torque term with $I^{\rm M}$ as well as a diffusion term with diffusion constant $D_{s(\bar s)}$.
Here, energy integrals for the electric field $I^{\rm EE}$ and the magnetic field $I^{\rm HE}$,
and magnetization integral $I^{M}$ are
defined, respectively, by
\begin{eqnarray}
I^{\rm EE}&&\equiv\oint d\boldsymbol M\times\boldsymbol E\cdot\hat{m}\nonumber\\
I^{\rm HE}&&\equiv\oint d\boldsymbol M\times\boldsymbol H_{\rm cons}\cdot\hat{m}=\oint H_{\rm cons}dM\nonumber\\
I^{\rm M}&&\equiv\oint d\boldsymbol M\times\boldsymbol M\cdot\hat{m}.
\end{eqnarray}
 On the assumption that the system is in steady state,
 the probability current density $\boldsymbol j_{\boldsymbol M}(t)$ vanishes.
 To get some insight in Eq.(\ref{curr}),
 we consider possible systems in steady state (i.e., $\boldsymbol  j_{\boldsymbol M}(t)=0$)
 and investigate the specific symmetry relation in each feasible case.
A steady state is reached if the condition $j_s(\varepsilon)=j_{\bar s}(\varepsilon)$ is satisfied.
After some straight forward algebra, we can write the relation corresponding to above condition 
$j_s(\varepsilon)=j_{\bar s}(\varepsilon)$ by
\begin{eqnarray}
&\mu_{0}I^{\rm HE}\bigg(\frac{D_{s}\nabla n_{s}-D_{\bar s}\nabla n_{\bar s}}{\nabla \varepsilon}\bigg)
=\nu I^{\rm EE}\big(n_{s}(\varepsilon)-n_{\bar s}(\varepsilon)\big)\nonumber\\
&-\frac{\gamma}{2}[(\alpha M-\frac{\gamma\hbar\tilde{g}^{s\bar s}}
{8\pi})I^{\rm HE}-J\hat{m}_{p}\cdot I^{\rm M}]
\big(n_{s}(\varepsilon)+n_{\bar s}(\varepsilon)\big)
\label{exge}.
\end{eqnarray}

 Equation (\ref{exge}) leads us to following feasible situations in the system.
First consider the case that up-spin and down-spin densities are the same.
 As long as the ratio of the work of the current induced torque to that 
 of the LL damping $\hat{m}_{p}\cdot I^{\rm M}/I^{\rm HE} $ is equal to 
 $\big(\alpha M-\frac{\gamma\hbar\tilde{g}^{s\bar s}}
{8\pi }\big)/J$
 and $I^{\rm HE}\neq 0$,
 the condition which guarantees the specific symmetry relation is 
\begin{eqnarray}\frac{\nabla n_{s}}{\nabla n_{\bar s}}\bigg\lvert_r=
 \frac{\nabla n_{s}}{\nabla n_{\bar s}}\bigg\lvert_{r'}=\frac{D_{\bar s}}{D_s}.\label{a1p}
 \end{eqnarray}
%Yet if $I^{\rm HE}=0$, the condition reduces to 
%\begin{eqnarray}\frac{\nabla n_{s}}{\nabla n_{\bar s}}\bigg\lvert_r=
% \frac{\nabla n_{s}}{\nabla n_{\bar s}}\bigg\lvert_{r'}=1.\label{a2p}
% \end{eqnarray}
 In this situation, 
 the specific symmetry relation is trivially satisfied 
 in normal region due to $D_s=D_{\bar s}$.
Another possible meaningful situation satisfying $j_s(\varepsilon)=j_{\bar s}(\varepsilon)$
 is that $n_{s}+n_{\bar s}=0$ resulted in $\nabla n_{s}+\nabla n_{\bar s}=0$.
 As long as $\nu I^{\rm EE}=0$ and $I^{\rm HE}=0$,
 the condition assuring the specific symmetry relation is given by 
 \begin{eqnarray}
 \frac{\nabla n_{s}}{\nabla n_{\bar s}}\bigg\lvert_r=
 \frac{\nabla n_{s}}{\nabla n_{\bar s}}\bigg\lvert_{r'}=-1.\label{a2p}
 \end{eqnarray}
 In this case, Eq.(\ref{a2p}) 
is in agreement with the 
previous condition \cite{ina} for the specific symmetry relation 
 in nonmetallic region of a homogeneous system with
 no space charge. 
 Even though the term containing the mixing conductance $\tilde{g}^{s\bar s}$ in Eq.(\ref{total}) 
 is neglected as the case of Slonczewski \cite{tse},
all the situations examined here satisfy the same conditions, Eqs.(\ref{a1p}) and (\ref{a2p}),
assuring the specific symmetry relation in the similar way with no terms of $\tilde{g}^{s\bar s}$.
As ever, these results are very useful to investigate spin related phenomena particularly 
in the normal metal region between two ferromagnets due to the condition $D_s=D_{\bar s}$.
 
 Regarding realistic systems such as quasi-two or three dimensional multilayer and 
 complicated electronic structures,
 corresponding spin-dependent scattering need to be considered.
Consequently, the neglected anisotropic terms and coordinate variables \cite{valet,Yu} have to be considered
 in electron distribution functions. 
 For this purpose, we examine the spin current density given by \cite{Ri}
\begin{equation}
\boldsymbol j_{s(\bar s)}=-e(\frac{m}{h})^3\int d^3v
\boldsymbol v g^{s(\bar s)}_{\boldsymbol v}(\varepsilon,\boldsymbol r)\label{cur}
\end{equation}
in terms of deviation distribution function 
$g^{s(\bar s)}_{\boldsymbol v}(\varepsilon,\boldsymbol r)$
corresponding to the energy $\varepsilon$ and velocity $v$.
A possible solution is given by 
\begin{eqnarray}
g^{s(\bar s)}_{v}(\varepsilon,\boldsymbol r)
=e\boldsymbol E\frac{\partial f_0(\varepsilon)}{\partial\varepsilon}
\lambda^{s(\bar s)}_{\rm eff}(\hat{\boldsymbol v},\boldsymbol r)
\label{dev}
\end{eqnarray}
for a system with all diffusive boundary effects driven by an 
applied electric field. It is valid 
in systems of various dimensions  
without neglecting anisotropic terms.
Here, the effective mean free path is given by \cite{Ri}
$\lambda^{s(\bar s)}_{\rm eff}(\vartheta,\boldsymbol r)=\lambda^{s(\bar s)}(\vartheta)
\bigg[1-{\rm exp}\big(\frac{-\lvert\boldsymbol r_0-\boldsymbol r\lvert}{\lambda^{s(\bar s)}(\vartheta)}\big)\bigg]
$ corresponding to an electron with normalized velocity $\hat{v}$ passing through
a position $\boldsymbol r$ in the direction $\boldsymbol r_0-\boldsymbol r$ with 
$\boldsymbol r_0$ on the cylindrical boundary.
The intrinsically anisotropic mean free path $\lambda^{s(\bar s)}(\vartheta)$ is 
given by $\lambda^{s(\bar s)}_0[1-a^{s(\bar s)}{\rm cos}^2\vartheta-b^{s(\bar s)}{\rm cos}^4\vartheta]$
where $\vartheta$ is the angle between the magnetization $\boldsymbol M$ 
and velocity $\boldsymbol v$.
When we compare Eq.(\ref{dev}) with another solution \cite{sui},
\begin{eqnarray}
\hat{v}\cdot e\boldsymbol E\lambda^{s(\bar s)}_{\rm eff}(\vartheta,\boldsymbol r)
%\bigg[1-{\rm exp}\big(\frac{-\lvert\boldsymbol r_0-\boldsymbol r\lvert}{\lambda^{s(\bar s)}(\vartheta)}\big)\bigg]\nonumber\\
=\bar\mu-\mu^{s(\bar s)}(\boldsymbol r)+\hat{v}\cdot e\boldsymbol E \lambda^{s(\bar s)}_0\label{mu1}
\end{eqnarray}
where $\bar\mu$ is the equilibrium electrochemical potential.
Accordingly, the relation convincing the specific symmetry relation is given by 
\begin{eqnarray}
&&\frac{N_{s}(\varepsilon_{\rm F})}{N_{\bar s}(\varepsilon_{\rm F})}
\frac{\nabla\big\{\hat{v}\cdot e\boldsymbol E\big[1-\lambda^{s}_{\rm eff}(\vartheta,\boldsymbol r)\big]\big\}-e\boldsymbol E
}{\nabla\big\{\hat{v}\cdot e\boldsymbol E\big[1-
\lambda^{\bar s}_{\rm eff}(\vartheta,\boldsymbol r\big]\big\}-e\boldsymbol E}
\Bigg\lvert_{\boldsymbol r}\nonumber\\
&&\qquad\qquad\qquad\frac{\nabla n_s}{\nabla n_{\bar s}}\bigg\lvert_{\boldsymbol r}=
\frac{\nabla n_s}{\nabla n_{\bar s}}\bigg\lvert_{\boldsymbol r'}.
\label{anssr}
\end{eqnarray}
Equation(\ref{anssr}) is based on the relation given, in the presence of an electric field ($\boldsymbol E=-\nabla{\Phi}$),
 by \cite{Yu}
 \begin{eqnarray}
\nabla n_{s(\bar s)}=eN_{s(\bar s)}(\varepsilon_{\rm F})[\nabla\mu_{s(\bar s)}-e\boldsymbol E].\nonumber
\end{eqnarray}
Here $N_{s(\bar s)}(\varepsilon_{\rm F})$ is the spin-up (down) density of states 
at the Fermi level in a highly degenerate system.
In order to generalize Eq.(\ref{anssr}),
we compare $g^{s(\bar s)}_{\boldsymbol v}(\varepsilon,\boldsymbol r)$ given by Eq.(\ref{mu1}) with 
another solution \cite{sheng} obtained by using the path integral along $\boldsymbol v$.
The gradient of the electrochemical potential
is then obtained by 
\begin{eqnarray}
&&\nabla\mu_{s(\bar s)}(\boldsymbol r)=
-\nabla\big[\hat{S}_{s(\bar s)\beta}(\boldsymbol r,\boldsymbol r')\hat{g}_{\beta\gamma}(\boldsymbol v,\boldsymbol r')
\hat{S}^{\dag}_{\gamma s(\bar s)}(\boldsymbol r,\boldsymbol r')\big]\nonumber\\
&&-\nabla\int_{\Gamma(\boldsymbol r,\boldsymbol r')}d{\it l''}
\hat{S}_{s(\bar s)\beta}(\boldsymbol r,\boldsymbol r'')\hat{v}\cdot e\boldsymbol E(\boldsymbol r'')
\hat{S}^{\dag}_{\gamma s(\bar s)}(\boldsymbol r,\boldsymbol r'')\frac{\partial f_{0}}{\partial\varepsilon}\nonumber\\
&&+\nabla\big(\hat{v}\cdot e\boldsymbol E \lambda^{s(\bar s)}_0\big)
\label{path1}
\end{eqnarray}
where the spinor propagation factor is given by 
$\hat{S}(\boldsymbol r,\boldsymbol r')=P_{\boldsymbol r'\rightarrow\boldsymbol r}
{\rm exp}(-\frac{1}{2}\int_{\Gamma(\boldsymbol r,\boldsymbol r')}d{\it l''}\hat{\xi}(\boldsymbol r''))$
with the inverse mean-free path operator $\hat{\xi}(\boldsymbol r'')$.
Here, $P_{\boldsymbol r'\rightarrow\boldsymbol r}$ is the path ordering operator along 
$\Gamma(\boldsymbol r,\boldsymbol r')$ 
which stands for the oriented straight path 
that starts at point $\boldsymbol r'$
and ends up at $\boldsymbol r$.
 If we take the limiting case of sufficiently large value of $\lvert\boldsymbol r-\boldsymbol r'\lvert$, 
Eq.(\ref{path1}) is written by
\begin{eqnarray}
&\nabla\mu_{s(\bar s)}(\boldsymbol r)=
-\nabla\int_{\Gamma(\boldsymbol r,\boldsymbol r')}d{\it l''}\big[
\hat{S}_{s(\bar s)\beta}(\boldsymbol r,\boldsymbol r'')\hat{v}\cdot e\boldsymbol E(\boldsymbol r'')\nonumber\\
&\hat{S}^{\dag}_{\gamma s(\bar s)}(\boldsymbol r,\boldsymbol r'')\frac{\partial f_{0}}{\partial\varepsilon}\big]
+\nabla(\hat{v}\cdot e\boldsymbol E \lambda^{s(\bar s)}_0).
\label{path}
\end{eqnarray}
The first term on the right hand side in Eq.(\ref{path}) corresponds to $-\nabla\big(\hat{v}\cdot e\boldsymbol 
E\lambda^{s(\bar s)}_{\rm eff}(\hat{\boldsymbol v},\boldsymbol r)\big)$
in Eq.(\ref{anssr}) assuring the specific symmetry relation.
Hence the condition assuring the symmetry relation is written by
$\frac{N_{s}(\varepsilon_{\rm F})}{N_{\bar s}(\varepsilon_{\rm F})}
\frac{\nabla[\hat{v}\cdot e\boldsymbol E \lambda^{s}_0-\int_{\Gamma(\boldsymbol r,\boldsymbol r')}d{\it l''}
\hat{S}_{s\beta}(\boldsymbol r,\boldsymbol r'')\hat{v}\cdot e\boldsymbol E (\boldsymbol r'')
\hat{S}^{\dag}_{\gamma s}(\boldsymbol r,\boldsymbol r'')\frac{\partial f_{0}}{\partial\varepsilon}]-e\boldsymbol E
}{\nabla[\hat{v}\cdot e\boldsymbol E \lambda^{\bar s}_0-\int_{\Gamma(\boldsymbol r,\boldsymbol r')}d{\it l''}
\hat{S}_{\bar s\beta}(\boldsymbol r,\boldsymbol r'')\hat{v}\cdot e\boldsymbol E (\boldsymbol r'')
\hat{S}^{\dag}_{\gamma\bar s}(\boldsymbol r,\boldsymbol r'')\frac{\partial f_{0}}{\partial\varepsilon}]-e\boldsymbol E}
\Bigg\lvert_{\boldsymbol r}\nonumber\\
\qquad\qquad\qquad=\frac{\nabla n_s}{\nabla n_{\bar s}}\big\lvert_{\boldsymbol r}
=\frac{\nabla n_s}{\nabla n_{\bar s}}\big\lvert_{\boldsymbol r'}$
in agreement with Eq.(\ref{anssr}).\

Hence, Eq.(\ref{anssr}) and above related condition
can be used to account adequately for experimental results.
Once some quantities such as $\boldsymbol v$, $\boldsymbol E$ and the angle $\vartheta$ at distinct positions are known,
we can check the properties of XCK with ease.
Besides, we can determine the relevance of theoretical and experimental consequences,
particularly in the region satisfying the specific symmetry relation 
such as a homogeneous nonmagnetic system.
For example, in case of choosing correct factors except an inappropriate $\lambda^{s(\bar s)}_{\rm eff}$,   
Eq.(\ref{anssr}) is not satisfied in trivial situations and proper modification for 
components such as $a^{s(\bar s)}$ and $b^{s(\bar s)}$ of $\lambda^{s(\bar s)}_{\rm eff}$
would be done to be in agreement with the relation. 
Also experimental values can be investigated as well and be compared with 
the modified calculations.  

Here, we have shown the conditions confirming the specific symmetry relation 
in current soft magnetic layered systems.
First, the specific symmetry relation has been elaborated 
by defining XCK explicitly on the ``mixed scheme'' compared with our pervious work \cite{ina}.
Then the conditions are derived by
taking into account the field gradients of the magnetic moment
in addition to that of the electric moment.
At the same time, special attention is paid to the fact that when we consider
the neglected anisotropic term in Boltzmann equation,
it gives deviation from generally accepted
definition of the electrochemical potential.
For the first time, we have thus given the physical condition by dealing with 
deviation distribution function
suitable for complex electronic structures in addition to one dimensional systems.

In conclusion, we have proposed a new method to confirm the relevance of studies of the spin-related phenomena
in soft magnetic layered systems of various dimensions by  taking 
the physical condition $\frac{\bigtriangledown n_{\bar\sigma}}{\bigtriangledown n_{\sigma}}\big\lvert_{\boldsymbol r}
=\frac{\bigtriangledown n_{\bar\sigma}}{\bigtriangledown n_{\sigma}}\big\lvert_{\boldsymbol r'}$ into account.
Once some quantities in the condition are given, 
the properties of XCK can easily be checked theoretically and experimentally
particularly in the region 
where the situation for convincing the specific symmetry relation is known.
Furthermore, excellent agreement of theoretical and experimental results can be induced by modifying 
factors or experimental conditions in accord with the physical condition.
\

One of the authors (IY) acknowledges support in part by the KOSEF (Grant No. R14-2002-029-01002-0) and 
(KSY) by the Korea Research Foundation 
(Grant No. KRF-2006-005-J02804). 

\renewcommand{\refname}{\large{References}}

\end{document}